\documentclass{article}

\usepackage[margin=1.5in]{geometry}
\usepackage{graphicx} 
\usepackage{amsmath}
\usepackage{amssymb}
\usepackage{amsthm}
\usepackage{xcolor}
\usepackage{setspace}
\usepackage[utf8]{inputenc}
\usepackage[T1]{fontenc}
\usepackage{float}
\theoremstyle{definition}
\usepackage[numbers]{natbib}
\usepackage[colorlinks=true, linkcolor=blue, citecolor=blue, urlcolor=blue]{hyperref}
\newtheorem{example}{Example}

\title{Shrinkage-Based Regressions with Many Related Treatments}

\author{%
  Enes Dilber \thanks{Wayfair}%
  \and
  Colin Gray\thanks{Work done at Wayfair}%
}

\date{April 2025}

\begin{document}

\maketitle

\begin{abstract}
When using observational causal models, practitioners often want to disentangle the effects of many related, partially-overlapping treatments. Examples include estimating treatment effects of different marketing touchpoints, ordering different types of products, or signing up for different services. Common approaches that estimate separate treatment coefficients are too noisy for practical decision-making. We propose a computationally light model that uses a customized ridge regression to move between a heterogeneous and a homogenous model: it substantially reduces MSE for the effects of each individual sub-treatment while allowing us to easily reconstruct the effects of an aggregated treatment. We demonstrate the properties of this estimator in theory and simulation, and illustrate how it has unlocked targeted decision-making at Wayfair.
\end{abstract}

\onehalfspacing

\section{Introduction}

Observational causal models are increasingly popular in industry,  often under the header of ``double machine learning`` or ``causal meta-learners'' \citep{chern-dml, causal-ml-book}. In practice, these methods are often used to study a single treatment or a small set of related treatments. However, many practitioners face a broad set of related but potentially sparse treatments and want to disentangle their individual impacts. Examples include estimating treatment effects of different marketing touchpoints, ordering different types of products, or signing up for different services. We refer to these as \textbf{sub-treatments}. This creates a dilemma: focusing on a single treatment typically yields precise estimates without the necessary granularity, while seperately estimating all sub-treatments typically results in estimates that are too imprecise to guide practical decision-making.


We propose a middle-ground between these extremes. Our core idea is to construct a ``focal function'' representing the single-treatment estimator, then use a specialized ridge regression to shrink additive interaction terms towards this component. This selective shrinkage improves the mean squared error (MSE) of causal estimates for each sub-treatment, leaves estimates essentially intact when a sub-treatment has sufficient data, and allows us to reconstruct the aggregate treatment effects for any degree of penalization without re-estimating the model. The use of closed-form ridge regression makes our method fast and scalable. The mechanics of this ridge regression also apply to estimating conditional average treatment effects, and to settings outside of causal meta-learners. 

This work builds heavily on the theory of causal meta-learners \citep{panel-dml, chern-dml, causal-ml-book, dr-learner, x-learner, r-learner}, and a smaller set of papers discussing their application in real-world settings \citep{cate-selection}. It is closely related to shrinkage using hierarchical Bayesian models \citep{bda-3}. We emphasize that our methods are primarily built for low-MSE estimation rather than rank-ordering treatments, for which partially linear models are sometimes inappropriate \citep{lal-rankings}. However, to our knowledge the results below will also apply to propensity score reweighted models.

Section~\ref{sec:modeling} illustrates our model and its properties. Section~\ref{sec:results} demonstrates the models mechanics using simulated data, and shows how Wayfair has used this idea to unlock practical decision-making at a more granular level. Section~\ref{sec:concl} concludes.

\section{Modeling Approach}
\label{sec:modeling}

\subsection{Structural Equations}
Our starting point is a partially linear \textbf{single treatment model}. For each unit $i$, we denote a scalar outcome $Y_i$ with a binary treatment $D_i$, a vector of confounders $X_i$ whose components are known but whose functional form $f$ is unknown, and an unobserved orthogonal error term $\epsilon_i$. This yields the structural equation:

\begin{equation}
    Y_i = f(X_i) + \beta D_i + \epsilon_i     \label{eq:single}
\end{equation}

Often, practitioners want to consider $K$ distinct treatments within a given business domain. As a running example, we consider the impact a potential customer being served a marketing touchpoint ($D_i$) when there are $K$ different types of marketing touchpoints (TV, search, social media, etc). We call these $K$ \textbf{sub-treatments}, and denote them $(D_{1i},...,D_{Ki})$. This motivates a regression with $K$ distinct treatment effects:

\begin{equation}
    Y_i = f(X_i) + \sum_k \theta_k D_{ki} + \epsilon_i    \label{eq:multi}
\end{equation}

We propose a more general formulation that gives a middle ground between the single treatment case (which is coarse but precise) and the sub-treatment case (which is flexible but imprecise). Denoting $D'_{ki} \equiv max_k(D_{ki})$, we write:

\begin{equation*}
    Y_i =  f(X_i) + \beta_0 \hspace{1.5mm} max_k(D_{ki}) + \sum_k \beta_k D_{ki} + \epsilon_i  
\end{equation*}

\begin{equation}
    = f(X_i) + \beta_0 \hspace{1.5mm} D'_{ki} + \sum_k \beta_k D_{ki} + \epsilon_i  \label{eq:max}
\end{equation}

We use the shorthand $D'_{ki}$ to save on notation, and to demonstrate that many of our theoretical results hold for a broader class of aggregation functions. We call this the \textbf{focal function}.

\subsection{Flexibly Residualizing Covariates}

We operationalize this with a flexible functional form $f$ using causal meta-learners. Following the steps in \citep{r-learner}, we assume that $E[\epsilon_i|X_i]=0 \forall X_i$ and apply a \textbf{Robinson transformation}:

\begin{align*}
    Y_i &=  f(X_i) + \beta_0 D'_i + \sum_k \beta_k D_{ki} + \epsilon_i \\
    E[\epsilon_i | X_i] &= 0 = E[Y_i|X_i] - f(X_i) - \beta_0 E[D'_i|X_i] - \sum_k \beta_k E[D_{ki}|X_i] \\
    & \Rightarrow Y_i -  E[Y_i|X_i] = \beta_0 (D'_i - E[D'_i|X_i]) + \sum_k \beta_k (D_{ki} - E[D_{ki}|X_i])
\end{align*}
    
Following convention, we denote residualized quantities with tildes:

\begin{align}
    \tilde{Y}_i = \beta_0 \tilde{D}'_i + \sum_k \beta_k \tilde{D}_{ki} + \epsilon_i  \label{eq:tildes}
\end{align}

\subsection{Shrinking Towards the Focal Function}

Using formulation \eqref{eq:max} and converting to the estimating equation \eqref{eq:tildes} gives us relatively simple way to compromise between the precision of a single-treatment model and the generality of a sub-treatment model. In particular, we shrink the coefficients $\beta_k$ towards zero, but do \textbf{not} penalize $\beta_0$. This reduces the MSE of each sub-treatment estimate, but (we show below) lets us reconstruct the single-treatment estimator. 

A simple and computationally efficient way to operationalize this is via ridge regression:

\begin{equation}
\label{eq:ridge_loss}
    min_{\beta_0, \beta_k} \big( Y_i - \beta_0 \tilde{D}'_i - \sum_k \beta_k \tilde{D}_{ki})^2 + \lambda || \beta_k ||_2 
\end{equation}

After tuning $\lambda$ using a hold-out sample, we can generate point estimates using a simple closed-form solution. Letting $X \equiv [\tilde{D}', \tilde{D}_{1}, \dots, \tilde{D}_{K}]$:

\begin{align*}
\hat{\beta}^{\lambda} &= \left(X'X + \Lambda\right)^{-1} X'Y \\
\widehat{\mathrm{Cov}}(\hat{\beta}^{\lambda}) &= \hat{\sigma}^2 \cdot \left( X' X + \Lambda \right)^{-1} X' X \left( X' X + \Lambda \right)^{-1}%
\footnotemark \\
\hat{\sigma}^2 &= \frac{1}{n - p} \left\| Y - X \hat{\beta}^{\lambda} \right\|^2 \\
\Lambda &= \mathrm{diag}(0, \lambda, \lambda, \dots, \lambda)
\end{align*}

\footnotetext{This is the homoscedastic form of the sandwich covariance estimator. A heteroscedasticity-robust version replaces $\hat{\sigma}^2 X'X$ with $X' \hat{\Sigma} X$, where $\hat{\Sigma}$ is a diagonal matrix whose entries are the squared residuals $\hat{\varepsilon}_i^2$.}


\subsection{Reconstructing Unbiased Estimators}

In addition to giving us low-MSE estimates of heterogeneous treatment effects, our formulation lets us reconstruct the single-treatment estimator for \textit{any} level of penalization. We show this in a few steps.

\subsubsection{\textit{We converge to the single-treatment estimator as penalties increase.}}

When ridge regularization is applied only to the sub-treatment parameters $\beta_k$ for $k > 0$, these coefficients shrink toward zero as $\lambda \to \infty$. Consequently, the ridge solution converges to a univariate projection of $Y$ onto $\tilde{D}'$, and the unpenalized parameter $\beta_0$ approaches:

\begin{equation}
\lim_{\lambda \to \infty} \beta_0^\lambda = \tau_0 \equiv \frac{\mathbb{E}[Y \tilde{D}']}{\mathbb{E}[\tilde{D}'^2]},
\label{eq:tau0_Y}
\end{equation}

\subsubsection{\textit{We can reconstruct the single-treatment estimator for any finite penalty.}}

Even when shrinkage is only partial, we can reconstruct an estimate of $\tau_0$ from our estimands. To see how, we expand the numerator of \eqref{eq:tau0_Y}:
\begin{align*}
\mathbb{E}[Y \tilde{D}'] &= \mathbb{E}\left[(\beta_0 \tilde{D}' + \sum_{k=1}^K \beta_k \tilde{D}_k + \epsilon) \tilde{D}'\right] \\
&= \beta_0 \mathbb{E}[\tilde{D}'^2] + \sum_{k=1}^K \beta_k \mathbb{E}[\tilde{D}' \tilde{D}_k] + \mathbb{E}[\tilde{D}' \epsilon].
\end{align*}

Assuming the noise term is mean independent of $\tilde{D}'$, i.e., $\mathbb{E}[\tilde{D}' \epsilon] = 0$, we obtain the following closed-form expression:
\begin{equation}
\tau_0 = \beta_0 + \sum_{k=1}^K \beta_k \cdot \frac{\mathbb{E}[\tilde{D}' \tilde{D}_k]}{\mathbb{E}[\tilde{D}'^2]}.
\label{eq:tau0_closed}
\end{equation}

To build intuition around this quantity, consider the case when $f(X) = 0$ (no selection bias) and $D' \equiv max_k(D_k)$. Then \eqref{eq:tau0_closed} simplifies to the sum of conditional probabilities:

\begin{align}
& \frac{\mathbb{E}[D' D_k]}{\mathbb{E}[D'^2]} = \frac{Pr[D'=1 \hspace{1mm} \& \hspace{1mm} D_k=1]}{Pr[D'=1]} = Pr[D_k|D'=1] \\
\Rightarrow \hspace{2mm} &  \tau_0 = \beta_0 + \sum_{k=1}^K \beta_k Pr[D_k|D'=1]
\label{eq:tau0_binary}
\end{align}

\subsubsection{\textit{This reconstructed estimator is invariant to the degree of penalization.}}

In ridge regression, the estimated coefficients $\beta_k^\lambda$ generally vary with the regularization parameter $\lambda$, affecting the contributions of the penalized variables $\tilde{D}_k$ to the fitted outcome. However, the quantity $\tau_0$ in Equation~\eqref{eq:tau0_closed}, which corresponds to the projection of $Y$ onto the unpenalized regressor $\tilde{D}'$, remains invariant to $\lambda$.

To see why, recall that the ridge estimator solves the optimization problem in Equation~\eqref{eq:ridge_loss}. Taking the derivative of the objective with respect to $\beta_0$ yields:
\[
\frac{\partial}{\partial \beta_0} 
\mathbb{E}\left[ \left( Y - \beta_0 \tilde{D}' - \sum_{k=1}^K \beta_k \tilde{D}_k \right)^2 + \lambda \sum_{k=1}^K \beta_k^2 \right]
= -2 \, \mathbb{E}\left[ (Y - \beta_0 \tilde{D}' - \sum_{k=1}^K \beta_k \tilde{D}_k) \tilde{D}' \right].
\]
Setting this derivative to zero gives the following first-order condition. And we know ridge estimated coefficients $\beta_k^\lambda$ satisfies it:
\[
\mathbb{E}\left[ (Y - \beta_0^\lambda \tilde{D}' - \sum_{k=1}^K \beta_k^\lambda \tilde{D}_k) \tilde{D}' \right] = 0.
\]
Letting $Y^\lambda = \beta_0^\lambda \tilde{D}' + \sum_{k=1}^K \beta_k^\lambda \tilde{D}_k$ denote the ridge-predicted outcome, this condition becomes:
\[
\mathbb{E}\left[ (Y - Y^\lambda) \tilde{D}' \right] = 0,
\quad \Rightarrow \quad
\mathbb{E}\left[ Y^\lambda \tilde{D}' \right] = \mathbb{E}\left[ Y \tilde{D}' \right].
\]
It follows that the projection coefficient of $Y$ onto $\tilde{D}'$ is:
\[
\tau_0^\lambda = \frac{\mathbb{E}[Y^\lambda \tilde{D}']}{\mathbb{E}[\tilde{D}'^2]} = \frac{\mathbb{E}[Y \tilde{D}']}{\mathbb{E}[\tilde{D}'^2]} = \tau_0.
\]

Thus, even though the individual coefficients $\beta_k^\lambda$ vary with $\lambda$, the combined projection $\tau_0^\lambda$ remains fixed. This implies that we can accurately reconstruct the single-treatment estimator regardless of the regularization strength applied to the sub-treatment terms.

\section{Results}
\label{sec:results}
\subsection{Simulation}

To demonstrate the behavior of our method in a controlled setting, we simulate a data-generating process with six binary sub-treatments. These sub-treatments are unevenly distributed, with activation probabilities ranging from 5\% to 20\%, reflecting common real-world imbalances. The outcome depends on both a shared treatment effect, activated whenever any sub-treatment occurs, and additive effects from each sub-treatment.

Specifically, for each unit $i$, we draw $D_{1i}, \dots, D_{6i} \sim \text{Bernoulli}(p_k)$ independently, where:
\begin{equation*}
p = [0.2, 0.05, 0.2, 0.05, 0.2, 0.05]
\end{equation*}
We define the focal treatment $D'_i = \max_k(D_{ki})$ and generate outcomes according to:
\[
Y_i = \beta_0 \cdot D'_i + \sum_{k=1}^6 \beta_k D_{ki},
\]
with $\beta_0 = 5$ and sub-treatment coefficients $\boldsymbol{\beta_k} = [2, 2, 1, 1, -1, -1]$.\footnote{This simulation excludes confounders to keep the focus on the mechanics of the specialized ridge regression.}

Similar to the derivation of the single-treatment effect in Equation~\eqref{eq:tau0_binary}, we can derive the expected impact of a specific sub-treatment as:

\begin{equation}
\tau_j = \beta_j +\beta_0 + \sum_{k\neq j}^K \beta_k \cdot P(D_k=1\mid D_j=1).
\label{eq:tauj_binary}
\end{equation}

\begin{figure}[ht]
    \centering
    \includegraphics[width=0.95\textwidth]{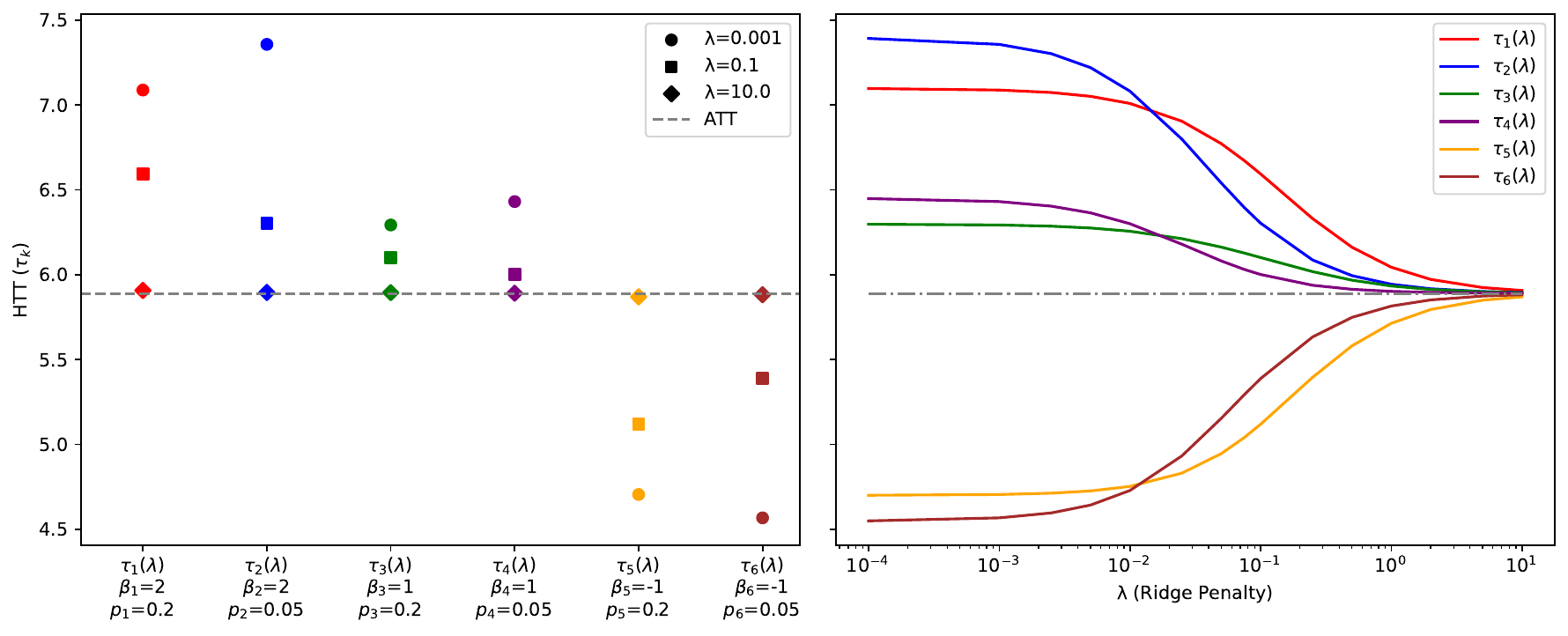}
    \caption{\footnotesize Sub-treatment effect estimates under varying ridge penalties. \textbf{Left:} Estimated heterogeneous treatment effects $\tau_k$ at three penalty levels. \textbf{Right:} Shrinkage paths showing convergence of each $\tau_k$ toward the single-treatment estimator (dashed line).}
    \label{fig:illustration}
\end{figure}

Figure~\ref{fig:illustration} illustrates how our estimator interpolates between high-variance heterogeneous treatment effects $\tau_k$ ($\lambda \approx 0$) and the shared average $\text{ATT}$ ($\lambda \rightarrow \infty$). Sparse sub-treatments shrink more aggressively, enabling stable and interpretable causal estimates even when sample sizes vary across treatments.

\begin{figure}[ht]
\centering
\includegraphics[width=0.85\textwidth]{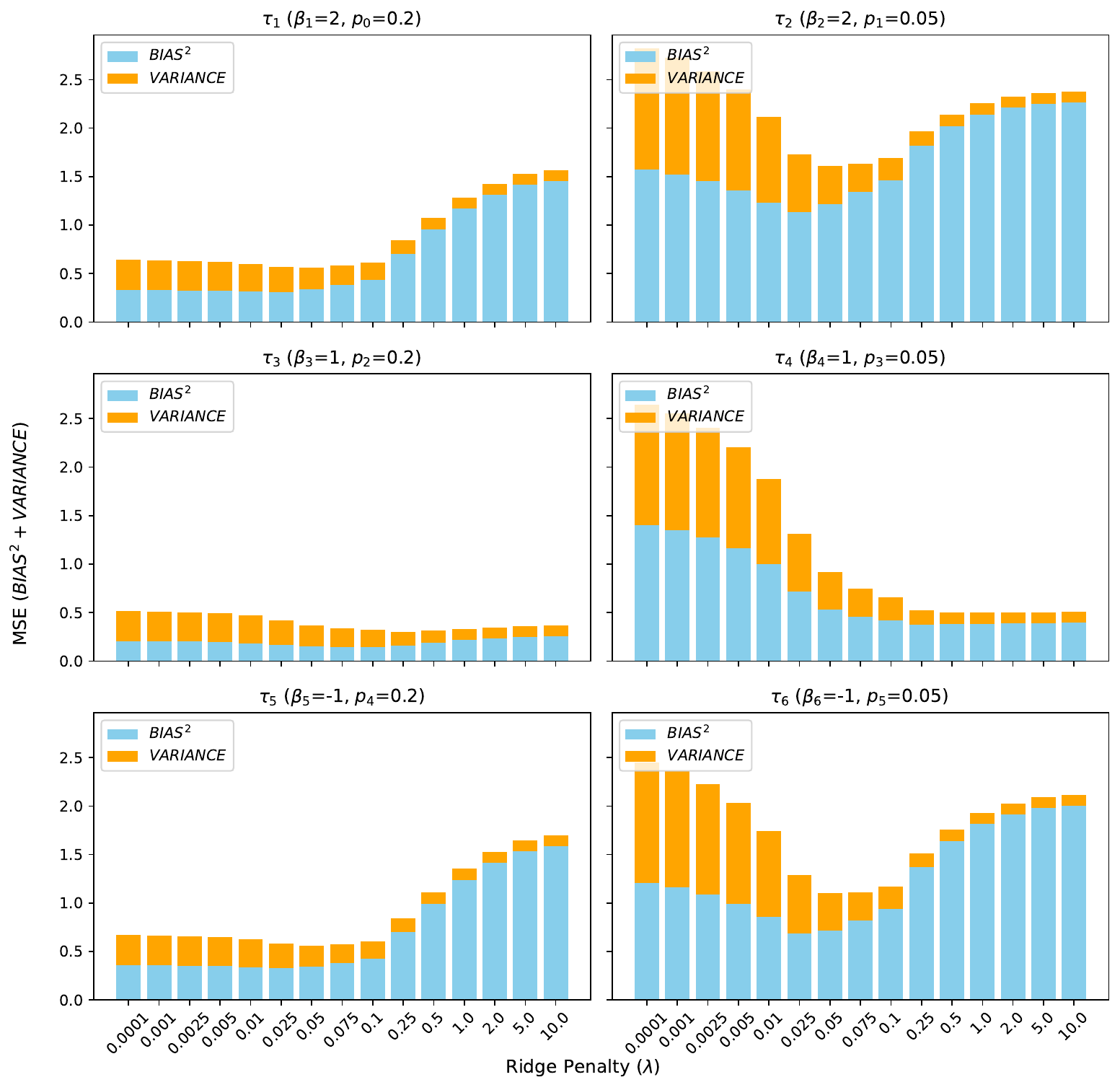}
\caption{\footnotesize Mean squared error (MSE) decomposition into squared bias and variance for each target parameter $\tau_k$ in Equation~\eqref{eq:tauj_binary} . Treatments with higher prevalence ($\tau_1$, $\tau_3$, and $\tau_5$) exhibit stable estimation across penalty values, while rarer treatments ($\tau_2$, $\tau_4$, and $\tau_6$) suffer high variance at low penalties.}
\label{fig:ridge_sim}
\end{figure}

Figure~\ref{fig:ridge_sim} confirms that the proposed estimator effectively trades off bias and variance across a range of regularization strengths. 

\subsection{Real-World Application at Wayfair}

We demonstrate our approach using real-world data from Wayfair, a large e-retailer that sells home goods ranging from small items (e.g., lamps) to large items (e.g., luxury beds). An important business decision in this context is whether certain product classes have an outsized long-term effect on customer spending. The shape of these effects is not obvious ex ante: it is possible that delivering small items reliably builds customer trust, or that delivering large, bulky items creates a more memorable positive experience.

We report anonymized heterogeneous sub-treatment effects from 53 product classes using our custom ridge regression method. The estimates are shown at three levels of regularization: no penalty ($\lambda = 0$), an optimally tuned penalty ($\lambda = 0.001$), and a large penalty ($\lambda = 100$), which approximates the projection of outcomes onto the focal treatment $\tilde{D}'$. As expected, increasing the penalty shrinks the class-specific estimates toward a shared baseline, similar to the behavior we observed in simulation.

\begin{figure}[ht]
    \centering
    \includegraphics[width=0.6\textwidth]{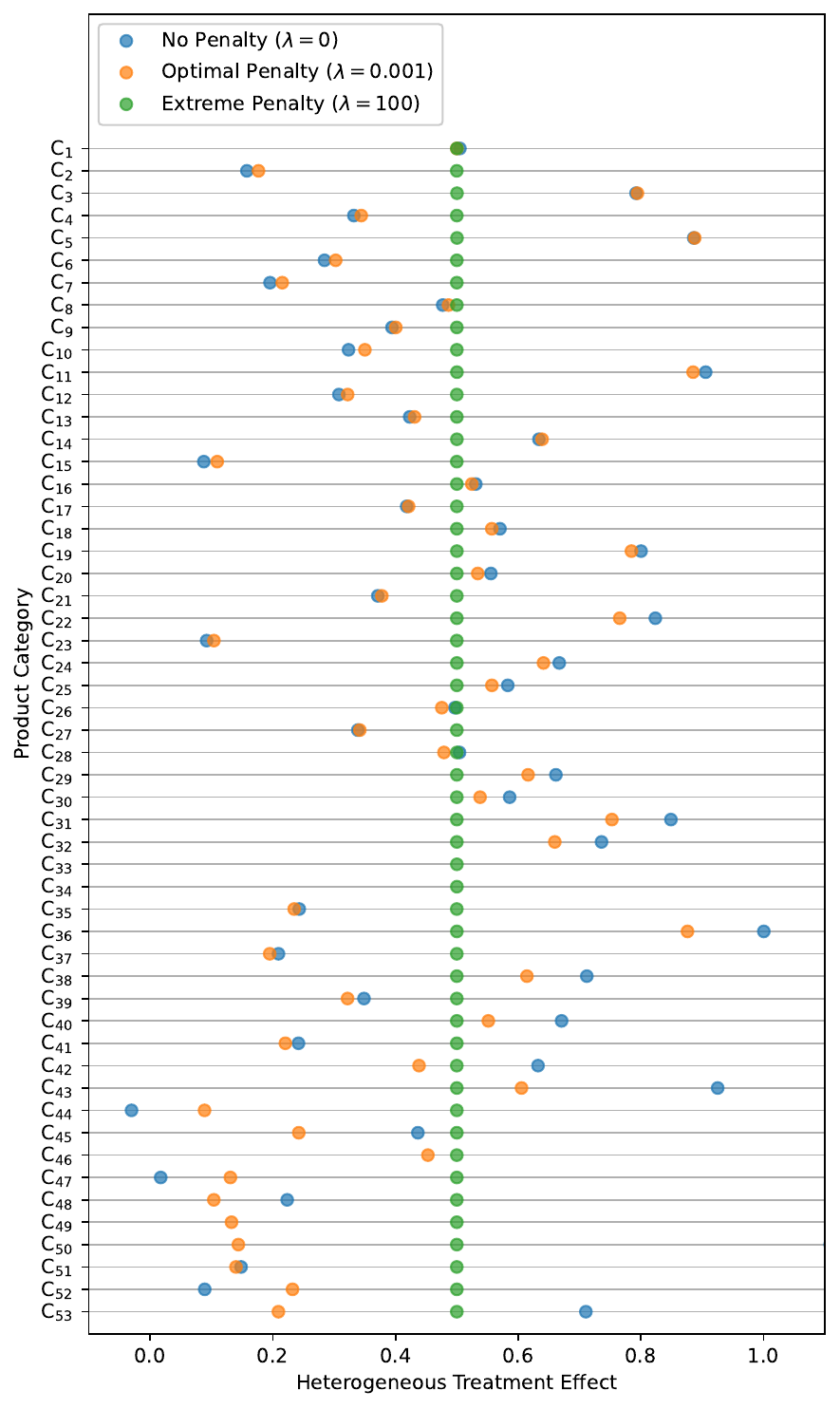}
    \caption{\footnotesize Estimated heterogeneous treatment effects for 53 anonymized product classes at three penalty levels: no penalty ($\lambda = 0$), optimal penalty ($\lambda = 0.001$), and extreme penalty ($\lambda = 100$). Classes are sorted by prevalence. High-prevalence categories (top, e.g., beds or rugs) exhibit lower variance and are less affected by regularization. Low-prevalence categories (bottom, e.g., flooring or pet) are more strongly shrunk toward the global average.}
    \label{fig:wayfair}
\end{figure}

Figure~\ref{fig:wayfair} highlights a key feature of our method: the ability to communicate treatment effects in a way that is robust to data sparsity. Because product classes are sorted by prevalence, we can visually confirm that higher-volume categories (e.g., beds, rugs) show stable estimates across penalties, while sparser classes (e.g., pet, flooring) are more sensitive to regularization. This effect is not an artifact of the method but a direct reflection of sample size–driven variance, which our approach dampens systematically.

We also tune the shrinkage parameter using a cross validation procedure and find that our method provides much more stable and interpretable effect estimates for business stakeholders. While the individual sub-treatment estimates are not unbiased, they enable more confident decision-making in settings where data volume varies dramatically across categories. We can further reconstruct the unbiased single-treatment estimates we would have obtained from extreme penalization (green) without re-estimating the model.

\section{Conclusion}
\label{sec:concl}

In practical business applications, we often use causal meta-learners to estimate separate effects of many partially-overlapping sub-treatments. Instead of choosing between coarse models with few treatments or very noisy models with every treatment, we propose a model that applies selective shrinkage towards a so-called focal function (the max across sub-treatments). In addition to providing lower-MSE estimates for each sub-treatment, we can easily reconstruct the single-treatment estimator in an internally-consistent and computationally efficient way. We have found this to be a flexible and pragmatic approach, which helps stakeholders make key business decisions with an appropriate degree of confidence.

\pagebreak

\bibliographystyle{chicago}
\bibliography{bib.bib}

\begin{thebibliography}{}

\bibitem[\protect\citeauthoryear{Athey and Imbens}{Athey and Imbens}{2025}]{panel-dml}
Athey, S. and G.~Imbens (2025).
\newblock {Identification of Average Treatment Effects in Nonparametric Panel Models}.
\newblock \url{https://arxiv.org/abs/2503.19873}.
\newblock arXiv preprint arXiv:2503.19873.

\bibitem[\protect\citeauthoryear{Chernozhukov, Chetverikov, Demirer, Duflo, Hansen, Newey, and Robins}{Chernozhukov et~al.}{2016}]{chern-dml}
Chernozhukov, V., D.~Chetverikov, M.~Demirer, E.~Duflo, C.~Hansen, W.~Newey, and J.~Robins (2016).
\newblock {Double/Debiased Machine Learning for Treatment and Causal Parameters}.
\newblock \url{https://arxiv.org/abs/1608.00060}.
\newblock arXiv preprint arXiv:1608.00060.

\bibitem[\protect\citeauthoryear{Chernozhukov, Hansen, Kallus, Spindler, and Syrgkanis}{Chernozhukov et~al.}{2024}]{causal-ml-book}
Chernozhukov, V., C.~Hansen, N.~Kallus, M.~Spindler, and V.~Syrgkanis (2024).
\newblock {Applied Causal Inference Powered by ML and AI}.
\newblock \url{https://arxiv.org/abs/2403.02467}.
\newblock arXiv preprint arXiv:2403.02467.

\bibitem[\protect\citeauthoryear{Gelman, Carlin, Stern, Dunson, Vehtari, and Rubin}{Gelman et~al.}{2013}]{bda-3}
Gelman, A., J.~Carlin, H.~Stern, D.~Dunson, A.~Vehtari, and D.~Rubin (2013).
\newblock {\em Bayesian Data Analysis (3rd ed.)}.
\newblock Chapman and Hall/CRC.

\bibitem[\protect\citeauthoryear{Kennedy}{Kennedy}{2020}]{dr-learner}
Kennedy, E.~H. (2020).
\newblock {Towards optimal doubly robust estimation of heterogeneous causal effects}.
\newblock \url{https://arxiv.org/abs/2004.14497}.
\newblock arXiv preprint arXiv:2004.14497.

\bibitem[\protect\citeauthoryear{Künzel, Sekhon, Bickel, and Yu}{Künzel et~al.}{2019}]{x-learner}
Künzel, S.~R., J.~S. Sekhon, P.~J. Bickel, and B.~Yu (2019).
\newblock {Metalearners for estimating heterogeneous treatment effects using machine learning}.
\newblock {\em Proceedings of the National Academy of Sciences\/}~{\em 116\/}(10), 4156--4165.

\bibitem[\protect\citeauthoryear{Lal}{Lal}{2024}]{lal-rankings}
Lal, A. (2024).
\newblock {Does Regression Produce Representative Causal Rankings?}
\newblock \url{https://arxiv.org/abs/2411.02675}.
\newblock arXiv preprint arXiv:2411.02675.

\bibitem[\protect\citeauthoryear{Mahajan, Mitliagkas, Neal, and Syrgkanis}{Mahajan et~al.}{2022}]{cate-selection}
Mahajan, D., I.~Mitliagkas, B.~Neal, and V.~Syrgkanis (2022).
\newblock {Empirical Analysis of Model Selection for Heterogeneous Causal Effect Estimation}.
\newblock \url{https://arxiv.org/abs/2211.01939}.
\newblock arXiv preprint arXiv:2211.01939.

\bibitem[\protect\citeauthoryear{Nie and Wager}{Nie and Wager}{2017}]{r-learner}
Nie, X. and S.~Wager (2017).
\newblock {Quasi-Oracle Estimation of Heterogeneous Treatment Effects}.
\newblock \url{https://arxiv.org/abs/1712.04912}.
\newblock arXiv preprint arXiv:1712.04912.

\end{thebibliography}

\pagebreak

\section{Appendix}
\label{sec:appdx}

\subsection[Interpreting tau\_0 under alternative focal functions]{Interpreting $\tau_0$ under alternative focal functions}
\label{appendix:tau0_cases}

Our main text shrinks granular estimates towards a "focal function" $f() = max_k(D_{ik})$. Other reasonable models of customer behavior could involve alternative focal functions, e.g. $f() = \sum_k D_{ik}$. Here, we show how $\tau_0$ can be interpreted as a valid treatment effect for both choices.

To be concrete, consider a scenario where we measure the effectiveness of a marketing campaign with two sub-treatments:

\begin{itemize}
    \item $D_1$: Seeing an ad on a website.
    \item $D_2$: Seeing the same ad via email.
\end{itemize}

Let $Y$ be the total spending after ad exposure and $D'$ be the focal function. Assuming a no-intercept model, the generative model is:

\[
Y = \beta_0 D' + \beta_1 D_1 + \beta_2 D_2+\epsilon.
\]

We focus on two focal functions.

\begin{example}[Maximum of binary sub-treatments]
Here we start with the most commonly used case in Wayfair.
\begin{equation*}
D' = \max(D_1, D_2),
\label{eq:dmax}
\end{equation*}

Which means the customer sees the ad in \textbf{at least one} of the two ways. 

\[
Y = \beta_0 D' + \beta_1 D_1 + \beta_2 D_2+\epsilon.
\]

Since $D_1$ and $D_2$ are binary variables, we can use the facts that (i) $E[D'D_k] = P(D_k=1, D'=1)$ and (ii) $E[D'^2] = P(D'=1)$ to write  Equation~\eqref{eq:tau0_closed} as:
\begin{equation*}
\tau_0=\beta_0+ \beta_1 P(D_1 = 1 | D' = 1) + \beta_2 P(D_2 = 1 | D' = 1)
\end{equation*}

This expression corresponds to the treatment effect given by the generative model above:

\[
E[Y\mid D' = 1] - E[Y\mid D' = 0].
\]
\end{example}

\begin{example}[Summation of binary sub-treatments]

Here we consider a model where number of times being exposed to the ads increases the outcome.

\begin{equation*}
D' = D_1 + D_2,
\label{eq:dsum}
\end{equation*}

Again, we are interested in the projected coefficient \( \tau_0 \) obtained by regressing \(Y\) onto \(D'\), which takes the form in Equation~\eqref{eq:tau0_closed}.

To relate this to interpretable quantities, we derive expressions for \( \mathbb{E}[Y \mid D' = d'] \) for \( d' = 1, 2 \). Because \( D' = D_1 + D_2 \), the values of \( (D_1, D_2) \) are restricted conditional on \( D' \):

\begin{itemize}
    \item When \( D' = 1 \), exactly one of \( D_1 \) or \( D_2 \) is equal to 1. Thus,
    \[
    \mathbb{E}[Y \mid D' = 1] = \beta_0 + \beta_1 \cdot \mathbb{E}[D_1 \mid D' = 1] + \beta_2 \cdot \mathbb{E}[D_2 \mid D' = 1].
    \]
    \item When \( D' = 2 \), both \( D_1 = 1 \) and \( D_2 = 1 \), so
    \[
    \mathbb{E}[Y \mid D' = 2] = 2\beta_0 + \beta_1 + \beta_2.
    \]
\end{itemize}

Next, we show that \( \tau_0 \) can be written as a convex combination of these conditional expectations. First, we calculate expectations \( \mathbb{E}[D'^2] \) and \( \mathbb{E}[D' D_k] \):

\[
\mathbb{E}[D'^2] = P(D' = 1) + 4 \cdot P(D' = 2),
\]
\[
\mathbb{E}[D' D_k] = P(D' = 1) \cdot \mathbb{E}[D_k \mid D' = 1] + 2 \cdot P(D' = 2) \cdot \mathbb{E}[D_k \mid D' = 2].
\]

Substituting these into the expression for \( \tau_0 \) in Equation~\eqref{eq:tau0_closed}, we get:

\begin{align*}
\tau_0
&= \beta_0 + \sum_{k=1}^2 \beta_k \cdot \frac{P(D' = 1) \cdot \mathbb{E}[D_k \mid D' = 1] + 2 P(D' = 2) \cdot \mathbb{E}[D_k \mid D' = 2]}{P(D' = 1) + 4 P(D' = 2)} \\
&= \frac{P(D' = 1)}{P(D' = 1) + 4 P(D' = 2)} \cdot \left( \beta_0 + \sum_{k=1}^2 \beta_k \cdot \mathbb{E}[D_k \mid D' = 1] \right) \\
&\quad + \frac{2 P(D' = 2)}{P(D' = 1) + 4 P(D' = 2)} \cdot \left( 2\beta_0 + \sum_{k=1}^2 \beta_k \cdot \mathbb{E}[D_k \mid D' = 2] \right).
\end{align*}

We now recognize the terms inside the parentheses as:

\[
\mathbb{E}[Y \mid D' = 1] = \beta_0 + \beta_1 \cdot \mathbb{E}[D_1 \mid D' = 1] + \beta_2 \cdot \mathbb{E}[D_2 \mid D' = 1],
\]
\[
\mathbb{E}[Y \mid D' = 2] = 2\beta_0 + \beta_1 + \beta_2,
\]
and since \( \mathbb{E}[D_k \mid D' = 2] = 1 \), the second term becomes consistent with our definition.

We conclude that:

\begin{equation}
\tau_0 = w_1 \cdot \mathbb{E}[Y \mid D' = 1] + w_2 \cdot \mathbb{E}[Y \mid D' = 2],
\end{equation}

where the weights are defined as:

\[
w_1 = \frac{P(D' = 1)}{\mathbb{E}[D'^2]}, \qquad
w_2 = \frac{2 \cdot P(D' = 2)}{\mathbb{E}[D'^2]},
\]

and satisfy \( w_1 + 2 w_2 = 1 \). This result shows that the projected coefficient \( \tau_0 \) is a convex combination of conditional means of \(Y\), weighted by the contribution of each value of \(D'\) to the total variation \( \mathbb{E}[D'^2] \).

\end{example}
\end{document}